\documentclass[prd,aps,superscriptaddress,showpacs,
preprintnumbers,floatfix,nofootinbib,twocolumn]{revtex4-1}

\usepackage{epsfig}
\usepackage{graphicx}
\usepackage{graphics}
\usepackage{pstricks,pstricks-add}
\usepackage{pst-plot}
\usepackage{epstopdf}
\usepackage{amsthm}
\usepackage{amsfonts}
\usepackage{amsmath}
\usepackage{amssymb}
\usepackage{bm}
\usepackage{rotating}
\usepackage{hyperref}
\usepackage{color}
\usepackage{revsymb}
\usepackage{bbold}
\usepackage{epsfig}

\newcommand{\dd}{\mathrm d}
\newcommand{\ex}{\mathrm e}
\newcommand{\lp}{\ensuremath{\left(}}
\newcommand{\rp}{\ensuremath{\right)}}
\newcommand{\lc}{\ensuremath{\left[}}
\newcommand{\rc}{\ensuremath{\right]}}
\newcommand{\lb}{\ensuremath{\left\lbrace}}
\newcommand{\rb}{\ensuremath{\right\rbrace}}
\newcommand{\MP}{M_\mathrm{_P}}

\newcommand{\GN}{G_\mathrm{_N}}

\definecolor{linkcolor}{rgb}{0.0,0.3,0.5}
\definecolor{venetianred}{rgb}{0.78, 0.03, 0.08}
\hypersetup{colorlinks=true,linkcolor=linkcolor,
citecolor=linkcolor,filecolor=linkcolor,urlcolor=linkcolor}

\begin{document}

\title{On the ghost-induced instability on de Sitter background}

\author{Patrick Peter}
\email{peter@iap.fr}
\affiliation{Institut d'Astrophysique de Paris (${\cal
		G}\mathbb{R}\varepsilon\mathbb{C}{\cal O}$), UMR 7095 CNRS}

\author{Filipe de O. Salles}
\email{salles@cbpf.br}
\affiliation{Centro Brasileiro de Pesquisas F\'isicas, 22290-180 URCA,
Rio de Janeiro (RJ), Brazil}

\author{Ilya L. Shapiro}
\email{shapiro@fisica.ufjf.br}
\affiliation{Departamento de F\'{\i}sica, ICE, Universidade Federal de
Juiz de Fora, 36036-330, MG, Brazil\\ Tomsk State Pedagogical University
and Tomsk State University, Tomsk, 634041, Russia}

\date{\today}

\begin{abstract}
\noindent It is known that the perturbative instability of tensor
excitations in higher derivative gravity may not take place if the
initial frequency of the gravitational waves are below the Planck
threshold. One can assume that this is a natural requirement if the
cosmological background is sufficiently mild, since in this case the
situation is qualitatively close to the free gravitational wave in flat
space. Here, we explore the opposite situation and consider the effect
of a very far from Minkowski radiation-dominated or de Sitter
cosmological background with a large Hubble rate, e.g., typical of an
inflationary period. It turns out that, then, for initial Planckian or
even trans-Planckian frequencies, the instability is rapidly suppressed
by the very fast expansion of the universe.
\end{abstract}



\maketitle

\section{Introduction}

The semiclassical approach to gravity is based on the equation
\begin{equation}
G_{\mu\nu} = R_{\mu\nu} -\frac12 Rg_{\mu\nu} =8\pi\GN \langle
T_{\mu\nu}\rangle, \label{semi}
\end{equation}
assuming that gravity itself is not quantized and the averaging in the
right hand side comes only from the quantum matter fields. Indeed, this
quantum average is quite nontrivial even in the vacuum case: it depends
on the curvature tensor and its derivatives, with a rich non-local
structure. It is well known that the renormalizable theory of matter
fields on a classical curved space-time background requires the action
of gravity to be an extension of General Relativity (GR)
\cite{Utiyama:1962sn} (see Refs.~\cite{Birrell:1982ix,Buchbinder:1992rb}
for an introduction to these topics and \cite{Shapiro:2008sf} for a more
recent review). The classical action of renormalizable semiclassical or
quantum gravity includes the usual Einstein-Hilbert term together with a
cosmological constant contribution,
\begin{equation}
S_{\mathrm{EH}}
=-\frac{1}{16\pi \GN}\int \dd^4 x\sqrt{-g}\lp R+2\Lambda \rp,
\label{EH}
\end{equation}
as well as the fourth derivative terms
\begin{equation}
S_{\mathrm{HD}} = \int \dd^4x \sqrt{-g}
\lp a_1C^2+a_2E+a_3{\Box}R+a_4R^2 \rp,
\label{HD}
\end{equation}
where $C^2=R_{\mu\nu\alpha\beta}^2 - 2 R_{\alpha\beta}^2 + \frac13 R^2$
is the square of the Weyl tensor and $E = R_{\mu\nu\alpha\beta}^2 - 4
R_{\alpha\beta}^2 + R^2$ is the integrand of the Gauss-Bonnet
topological term. All the terms of this vacuum action
\begin{equation}
S_{\mathrm{vac}}=S_{\mathrm{EH}}+S_{\mathrm{HD}}
\label{vacuum}
\end{equation}
belong to the gravitational action, but it is traditional to write the
field equations in the form of Eq.~(\ref{semi}) and to include the
higher derivative contributions to the right hand side. The same action
(\ref{vacuum}) leads to the simplest renormalizable theory of quantum
gravity \cite{Stelle:1976gc}.

Unfortunately, the very same terms (\ref{HD}) which provide
renormalizability also lead to a serious problem, since they produce
unphysical ghosts and hence induce instabilities of classical solutions.
In turn, trying to remove these ghosts from the spectrum renders the
gravitational $S$-matrix non-unitary
\cite{Stelle:1976gc,Buchbinder:1992rb}. This last fact is however not a
real problem if one restricts attention to semiclassical gravity, since
then the $S$-matrix of gravitons is of no relevance. Hence, studying the
question of ghosts essentially reduces to that of investigating possible
instabilities in the classical solutions.

It was recently stressed, in particular in the recent works by some of
the present authors and collaborators
\cite{Salles:2014rua,Cusin:2015oza}, that constructing a consistent
theory without higher derivative terms is not possible, because the
corresponding coefficients are logarithmically running. Let us briefly
elaborate on this issue. Imagine we decide to avoid ghosts and require
that the coefficient $a_1$ in the action \eqref{HD} is exactly zero. Then
the loops of matter fields produce the same term with divergent
coefficient. One can subtract the unphysical local divergent term, but
in the UV limit there will be a logarithmic form factor, such that the
relevant quantum contribution has the form
\begin{equation}
\int \dd^4x\,\sqrt{-g}\,C_{\alpha\beta\rho\sigma}
\ln\left( \frac{\Box}{\mu^2}\right) C^{\alpha\beta\rho\sigma}.
\label{ff}
\end{equation}

At low energies the non-local form factor is suppressed by the
decoupling effect \cite{Gorbar:2002pw}, and in principle one can use the
described scheme of renormalization to avoid higher derivatives.
However, in the UV there is no decoupling, and since the logarithm
function in Eq.~(\ref{ff}) is slowly varying, we arrive, effectively, at
the action (\ref{HD}) with the coefficient $a_1$ which is defined by the
$\beta$-function. In the formalism of anomaly-induced effective action,
the same effect is achieved in a very elegant way, as we will describe
in Sec.~\ref{s2}. This effective action represents a local version of
the renormalization-group improved classical action
\cite{Shapiro:2008sf}. All in all, we can see that the renormalization
group running of some parameter indicates that this parameter cannot be
fixed to be zero, at least in the UV. In semiclassical gravity, this is
exactly what happens with $a_1$, and this makes the discussion of ghosts
really pertinent. Let us note that this feature constitutes a
fundamental difference with, e.g., the higher derivative terms generated
as quantum corrections in QED, since in this case the coefficients of
the higher derivative terms do not run, and therefore can be safely
regarded as small corrections, naturally providing a way to avoid the
runaway solutions. At the moment, no consistent scheme of this kind
exists for gravity.

There is currently no satisfying solution to the conflict between
renormalizability and ghost-induced instabilities, but there are
remarkable facts concerning unitarity of renormalizable or
superenormalizable \cite{Asorey:1996hz} quantum gravity which are worth
mentioning. From the non-perturbative analysis, there appears to be a
chance that quantum (or semiclassical) contributions to the propagator
of gravitons may split the ghost pole into a pair of complex conjugate
poles. As a result, unitarity of the $S$-matrix can be restored
\cite{Tomboulis:1977jk,Tomboulis:1980bs,Tomboulis:1983sw,
Salam:1978fd,Antoniadis:1986tu}. However, in order to know whether this
really happens, one needs a full-nonperturbative knowledge of the
dressed propagator of gravitational perturbations
\cite{Johnston:1987ue}. Recently, it was shown that the restoration of
unitarity can be achieved by introducing a special UV completion to the
action \eqref{HD} in the form of six-derivative terms
\cite{Asorey:1996hz} already at the tree-level.  In this case, the
unitarity of the gravitational $S$-matrix is guaranteed provided these
terms ensure that the massive poles appear in complex conjugate pairs
\cite{Modesto:2015ozb,Modesto:2016ofr}.

An even more dramatic effect can be achieved by assuming a non-local
form factor which can make the tree-level theory to be free from all
ghost-like poles. The original construction of this kind was suggested
in the framework of string theory \cite{Tseytlin:1995uq}, but it
recently gained popularity as a proposal for an unusual quantum gravity
setup \cite{Tomboulis:1997gg} (see further developments in
\cite{Modesto:2011kw,Modesto:2014lga}). Unfortunately, in the latter
case, ghosts always come back when loop corrections are taken into
account \cite{Shapiro:2015uxa}. All this concerns the unitarity of the
$S$-matrix, while the issue of stability in this kind of theories with
or without ghosts is not explored yet. One can consider the situation
from a slightly different perspective. Since the ghost mass is typically
of the Planck order of magnitude \cite{Accioly:2016etf,Accioly:2016qeb},
the instability coming from ghosts implies the possibility to accumulate
gravitons with Planck energy density in a small volume of space, where
the ghost particle can be created from vacuum. The stability of such a
theory with ghosts therefore implies that there exists some mechanism
thanks to which this accumulation is prevented.  We do not know how this
mechanism works \cite{Dvali:2010ue,Dvali:2011aa}, but one can imagine
that it can be related to the non-local form factor in the gravitational
action, e.g., be similar to that which prevents the Newtonian or black
hole singularity. Such a form factor can also be local (polynomial in
momenta) \cite{Modesto:2014eta}. Then the results of
Ref.~\cite{Salles:2014rua} show that the instability does not occur
provided the cosmological background is slowly varying, i.e. is such
that the energies involved (inverse time, wavenumber,...) are all
negligible with respect to the Planck scale, and that the initial
frequencies of the perturbations are also sub-Planckian. Mathematically,
linear stability on a given background is, for a small initial
amplitude, a sufficient condition of stability at higher (finite) level
in the perturbation theory.

In what follows, we shall concentrate on the stability problem in the
simplest minimalist model  (\ref{vacuum}), and leave more complicated
models mentioned above for future work. We continue along the line of
our previous work \cite{Salles:2014rua} and discuss the situation in the
basic fourth-derivative theory [i.e., actions \eqref{EH} and
\eqref{HD}], together with quantum corrections, which are  taken into
account by integrating the conformal anomaly. The method of deriving the
gravitational wave equation in the theory with quantum corrections is
technically simple \cite{Fabris:2011qq}, and the results are
qualitatively equivalent to those previously obtained by Starobinsky for
the gravitational wave equation on the de Sitter background
\cite{Starobinsky:1979ty,Starobinsky:1981zc,Starobinsky:1983zz}.

The main point of the present work is twofold. First, we elaborate
further on the result of \cite{Salles:2014rua} (see also the previous
papers \cite{Fabris:2011qq,Fabris:2000gz} and the short review
\cite{Shapiro:2014fsa}), stating that there is no amplification of the
gravitational waves in the higher derivative gravity, including with
quantum corrections, if the cosmological background is relatively mild
and (most important) if the initial frequency of the gravitational
perturbation is well below the Planck scale. This was in fact well-known
from the previous studies on de Sitter background in
\cite{Starobinsky:1979ty,Starobinsky:1981zc,Starobinsky:1983zz} and more
recently in \cite{Hawking:2000bb} and \cite{Pelinson:2002ef}. By no
means can it be seen as a surprise that the same non-amplification takes
place for the radiation and matter-dominated backgrounds. However, it
was found in \cite{Salles:2014rua} that there actually {\it is} a very
strong amplification of the gravitational waves starting from the
Planck-order frequencies. The second point of the present work is made
once we assume that the frequencies of perturbations above the Planck
scale occurs independent on the type of cosmology (fluid domination or
de Sitter,...), but only for values of the Hubble scale of the
background that are comparable with the Planck scale $\MP=\GN^{-1/2}$.

It is natural to think that the two requirements, namely that of
small typical energy of metric perturbations and of a slowly-varying
background, are not independent since there can be energy exchange
between the background and perturbations. For linear perturbations,
this is simply a restating of the obvious fact that the background
can affect perturbations. In what follows we shall explore the effect
of a strong background on the dynamics of  high-frequency tensor 
modes of metric perturbations.

The paper is organized as follows. In Sect. II, we briefly introduce the
effective equations induced by the fourth-derivative classical terms
(\ref{HD}) and by the quantum corrections related to the conformal
anomaly; we summarize the equations for both the background and the
tensor perturbations. Sect. III provides an analysis of the initial
conditions that are necessary to solve our equations while Sect. IV
contains our results; we present a numerical analysis as well as some
qualitative discussion of the stability, including the most important
cases of Planck-order frequencies and fast-varying background. Finally,
we draw our conclusions  in the final Sect. V and discuss some of the
unsolved issues in our approach.

\section{Effective equations induced by anomaly}
\label{s2}

The analysis is performed for the higher derivative action (\ref{vacuum})
and the same quantum corrections which were already discussed in
Refs.~\cite{Fabris:2011qq,Salles:2014rua,Shapiro:2014fsa}; a complete
and detailed derivation/discussion of the relevant equations for both
background and perturbations being available in these references, we
shall heavily rely on those to present a mere brief introduction to the
matter before going on to our point.

Let us briefly review the anomaly-induced effective action and the
corresponding equation for metric perturbations. The anomalous
trace of the energy momentum tensor is well-known (see, e.g.,
\cite{Birrell:1982ix}) and given by
\begin{equation}
\langle T_\mu^\mu \rangle = -\lp \omega C^2 + bE + c\Box R \rp,
\label{tracean}
\end{equation}
where the coefficients $\omega$, $b$ and $c$ depend on the number of
active quantum fields of different spins. Let us note that the
expression (\ref{tracean}) and therefore the related physical results
can be regarded as being non-perturbative with respect to the loop
expansion. In renormalizable semiclassical theories, the general
structure of the trace anomaly (\ref{tracean}) is supposed to be the
same at the one-loop level and beyond. The main difference is that at
higher loops the $\beta$-functions $\omega$, $b$, $c$ become power
series in the coupling constants.

From \eqref{tracean}, one derives the anomaly-induced effective
action $\bar{\Gamma}_\mathrm{ind}$, obtained by integrating the
equation
\begin{equation}
\frac{2}{\sqrt{-g}} g_{\mu\nu}
\frac{\delta {\bar \Gamma}_{\mathrm{ind}}}{\delta g_{\mu\nu}}
=- \langle T_\mu^\mu \rangle =\omega C^2 + bE + c\Box R.
\label{mainequation}
\end{equation}
The covariant and local solution of \eqref{mainequation} has been
proposed in Refs.~\cite{Riegert:1984kt,Fradkin:1983tg}, and the 
most complete form involving two auxiliary fields has been found 
in \cite{Shapiro:1994ww} (see also an equivalent form constructed 
independently in \cite{Mazur:2001aa}). The corresponding 
expression reads
\begin{widetext}
\begin{equation}
\bar{\Gamma}_\mathrm{ind}=S_{\mathrm{c}}[g_{\mu\nu}] +
\int \dd^4x \sqrt{-g}  \lb 
- \frac{k_3}{12} R^2+ \frac12 \phi\Delta_4\phi + \phi\lc k_1 C^2 +
k_2\lp E - \frac{2}{3}\Box R\rp\rc - \frac12\psi\Delta_4\psi + l_1
C^2 \psi \rb,
\label{tota}
\end{equation}
\end{widetext}
where $\Delta_4=\Box^2+2R^{\mu\nu}\nabla_\mu \nabla_\nu
-\frac23\,R\Box+\frac13\,R^{;\mu}\nabla_\mu$ is the covariant conformal
fourth order operator  \cite{Riegert:1984kt,Fradkin:1983tg} and the
coefficients are given in terms of those of \eqref{tracean} through
\begin{equation}
k_1 = - \frac{\omega}{2\sqrt{-b}}
 ,\quad
k_2=\frac{\sqrt{-b}}{2}
 ,\quad
k_3 = c + \frac{2}{3}b
 ,\quad
l_1=\frac{\omega}{2\sqrt{-b}}.
\end{equation}
Furthermore, the relevant $\beta$-functions depend on the numbers of
real scalar degrees of freedom $N_0$, four-component spinor fermions
$N_{1/2}$ and vector fields $N_1$ in the underlying particle physics
model, leading to
\begin{equation}
\left(
\begin{array}{c}
\omega  \\
b       \\
c        \\
\end{array}
\right)
=
\frac{1}{360 (4\pi)^2}
\left(
\begin{array}{ccc}
3 N_0 + 18 N_{1/2} + 36 N_1
\\
-  N_0 - 11 N_{1/2} - 62 N_1
\\
2 N_0 + 12 N_{1/2} - 36 N_1    \\
\end{array}
\right).
\label{wbc}
\end{equation}
For the Minimal Standard Model (MSM) of Particle Physics, 
based on the SU(3)$\times$SU(2)$\times$U(1) gauge group, 
with 8 gluons, 3 intermediate vectors $W^\pm$ and $Z^0$ and 
the photon, this gives $N_1=12$, the Higgs SU(2) doublet leads 
to $N_0=4$, and the 3 lepton and quark SU(2) doublets, assuming 
the neutrino to be massive, imply $N_{1/2} = 24$ (and hence $c<0$).

Finally, the action $S_{\mathrm{c}}[g_{\mu\nu}]$ is an unknown 
conformal invariant functional which can be seen as an integration 
constant of Eq.~(\ref{mainequation}). For the background cosmological 
solutions, it is irrelevant. Moreover, as discussed in
Refs.~\cite{Balbinot:1999ri,Balbinot:1999vg,Shapiro:2008sf}, there 
are also very convincing reasons to disregard it in many other 
situations, an attitude we shall adopt from now on.

The cosmological background solution in the theory based on the action
$S_\mathrm{vac}+\Gamma_\mathrm{ind}$ can be explored by assuming
\begin{equation}
g_{\mu\nu} = a^2(\eta) {\bar g}_{\mu\nu}
 ,\quad
{\bar g}_{\mu\nu} = \eta_{\mu\nu}
 ,\quad
a(\eta) = \ex^{\sigma(\eta)} ,
\label{back}
\end{equation}
where $\eta$ is the conformal time defined through $\dd t =
a(\eta)\dd\eta$. In this case, the equations for the auxiliary fields
$\phi$ and $\psi$ reduce to
\begin{equation}
{\Box}^2 \lp \phi
+ 8\pi\sqrt{-b} \sigma \rp = 0\ \ \ \hbox{and} \ \ \
{\Box}^2 \psi = 0 .
\label{uravnilovki}
\end{equation}
Here $\Box = \partial_t^2 - \bm{\nabla}^2$ is the d'Alembertian operator
constructed with the flat metric. The solutions of (\ref{uravnilovki})
can be cast in the form
\begin{equation}
\phi = - 8\pi\sqrt{-b} \sigma + \phi_0 ,
\qquad
\psi =  \psi_0 .
\label{reshen}
\end{equation}
where both $ \phi_0$ and $\psi_0 $ are general solutions of the
homogeneous equation $ {\Box}^2 f =0$ corresponding to the 
fiducial metric ${\bar g}_{\mu\nu}$. In the cosmological
case \eqref{back}, for which $\sigma=\ln a$, the time
derivatives are simply given in terms of the Hubble growth
rate, namely
$$\dot{\phi}=-8\pi\sqrt{-b}\,H,\quad
\ddot{\phi}=-8\pi\sqrt{-b}\,{\dot H},\quad
\stackrel{...}{\phi}=-8\pi\sqrt{-b}\,{\ddot H},$$
and so on.

Replacing these solutions back into the action $\Gamma_\mathrm{ind}
+S_\mathrm{EH} +S_\mathrm{HD}$ and taking variations with respect to
$\sigma$, one arrives at the equation (more details are available in
Ref.~\cite{Pelinson:2002ef}) 
\begin{align}
&& \frac{a^{\lp\textsc{\tiny
IV}\rp}}{a} +\frac{{3\dot{a}} {a^{\lp\textsc{\tiny III}\rp}}}{a^2}
+\frac{\ddot{a}^2}{a^2} -\lp 5+\frac{4b}{c}\rp \frac{\ddot{a}
\dot{a}^2}{a^3} \nonumber \\ && -\frac{\MP^2}{8\pi c} \lp
\frac{\ddot{a}}{a}+ \frac{\dot{a}^2}{a^2} -\frac23\Lambda\rp = 0 ,
\label{foe}
\end{align}
where $t$ is the physical time and we have used that $\ex^\sigma
\dd\eta=a \dd\eta=\dd t$; in \eqref{foe} and the following, a dot stands
for a time derivative and we wrote $f^{\lp\textsc{\tiny
III}\rp}=\partial^3 f/\partial t^3$ and $f^{\lp\textsc{\tiny
IV}\rp}=\partial^4 f/\partial t^4$. The last equation does not take into
account matter and space curvature. This is easily justified by the
extremely fast expansion of the universe in the inflationary epoch which
we intend to describe. It should be noted that Eq.~\eqref{foe} depends
only on $b$ and $c$, and not on $\omega$, the latter entering only
through the conformal invariant Weyl tensor, which cannot contribute to
the conformal Friedman-Lema\^{\i}tre-Robertson-Walker (FLRW) solution
\eqref{back}. Similarly, this equation can depend neither on $a_1$, nor
on $a_2$ and $a_3$ (surface term), and we assume for now on that $a_4=0$
to ensure a conformal invariant initial theory.

A detailed discussion of the general solution of this equation can be
found in Refs.~\cite{Antoniadis:1986tu,Starobinsky:1980te,
Starobinsky:1981vz,Shapiro:2001rh}, and in particular the inflationary
solutions in the presence of a cosmological constant were obtained in
\cite{Pelinson:2002ef}. These two important particular solutions are
both for the de Sitter scale factor $a(t)  =  a_0 \ex^{Ht}$, with
\begin{equation}
H =  \frac{\,\,\MP}{\sqrt{-32\pi b}} \Bigg(1\pm
\sqrt{1+\frac{64\pi b}{3}\frac{\Lambda }{\MP^2}} \Bigg)^{1/2} .
\label{H}
\end{equation}
Since the cosmological constant satisfies the condition $ \Lambda \ll
\MP^2$, one gets, assuming $b\not\gg 1$, upon expanding, the following
two vastly different values of  $H$,
\begin{eqnarray}
H_\mathrm{classical} \approx \sqrt{\frac{\Lambda }{3}}
\quad
{\rm and}
\quad
H_0 \approx \frac{\,\,\MP}{\sqrt{-16\pi b}}.
\label{HH}
\end{eqnarray}
where $H_\mathrm{classical}$ corresponds to the de Sitter space 
without quantum corrections, and the value $H_0$ gives the 
exponential solution of Starobinsky \cite{Starobinsky:1980te}. 

According to (\ref{wbc}), the constant $b$ is always negative,
irrespective of the actual numbers of scalar, fermionic and vectorial
degrees of freedom, whereas that of  $c$ explicitly depends on the
particle content of the theory together with the finite value of the
$R^2$ term introduced into the classical action $S_\mathrm{HD}$, as
shown in Eq.~(\ref{HD}).

It turns out that the solution \eqref{HH} is stable with respect to
variations of the initial data for $\sigma(t)$
\cite{Starobinsky:1980te,Pelinson:2002ef}, provided the parameters 
of the underlying quantum theory satisfy the condition $b/c < 0$, 
which translates, since $b<0$, into the condition $c>0$, i.e., given 
\eqref{wbc}, to the relation 
\begin{equation}
N_1 <
\frac13 N_{1/2} + \frac{1}{18} N_0.
\label{const}
\end{equation}
This constraint is not satisfied for the standard model, as discussed
above [see Eq.~\eqref{wbc}]. However, there are many reasons to 
suspect the SM to not be the end of the story and many extensions 
have been proposed, having many more degrees of freedom and for which 
the inequality (\ref{const}) can readily be satisfied. For instance, a 
minimal supersymmetric extension of the MSM (MSSM), demands 
$N_1=12$, $N_{1/2}=32$ and $N_0=104$, which implies $c>0$ as 
required for inflation to initiate in the stable phase 
\cite{Shapiro:2001rd}. The same sign of $c$ is expected for any 
version of phenomenologically acceptable supersymmetric extension 
of the Standard Model.  It is worth mentioning that the transition to 
unstable Starobinsky inflation has been described in 
\cite{Shapiro:2001rd,Shapiro:2001rh,Pelinson:2002ef} and more 
recently in \cite{Netto:2015cba}. 

The advantage of having an inflation that is stable, i.e. with a
particle content satisfying \eqref{const}, is that the inflation phase
occurs independently of the initial data. After the initial singularity
(or whatever replaces it when the theory is smoothed at the relevant
scale), when the Universe starts expanding and the typical energy
decreases below the Planck scale, one can envisage some transition
(stemming from string theory or whatever actually describes the physics
at this scale) below which the effective quantum field theory is an
adequate description, and the anomaly-induced model applies. The
specificity of the case above is that it does not need any fine
tuning for the initial value of either $a(t)$ or its time derivatives, the
only requirement being that the condition (\ref{const}) holds.

\section{Tensor perturbations}

Expanding the FLRW metric and restricting attention to the tensor mode
only yields
\begin{equation}
\dd s^2=a^2(\eta)\lc \dd\eta^2 -  \lp\gamma_{ij} + h_{ij}\rp \dd x^i
\dd x^j\rc,
\label{pertmet}
\end{equation}
where the perturbation $h_{ij}$ is traceless ($\gamma^{ij} h_{ij}=0$)
and transverse ($\gamma^{ik}\partial_k h_{ij}=0$) \cite{PPJPU2013}.
\begin{widetext}
\noindent These two conditions ensure that we are actually dealing with
the tensor component of $h_{ij}$, leaving the scalar and vector parts
out. The relevant two degrees of freedom describing these gravitational
waves correspond to the well-known $(+)$ and $(\times)$ polarization
states. For the sake of simplicity, we assume in what follows the
background metric to be flat, so we fix $\gamma_{ij}=\delta_{ij}$.
	
Our gravity theory with anomaly-induced corrections is described by a
Lagrangian density comprising all the terms in Eqs.~\eqref{vacuum} and
\eqref{tota}. Gathering all similar terms, it can be rewritten in the
form
\begin{eqnarray}
\label{1}
{\cal L}
= \sum^{5}_{s = 0} f_{\mathrm{s}} {\cal L}_{\mathrm{s}}
= f_{0} R
+ f_1 R^2_{\alpha\beta\mu\nu}
+ f_2R^2_{\alpha\beta}
+ f_3 R^{2}
+ f_4\phi\Box R
+ \frac12 \phi \Delta \phi,
\end{eqnarray}
where the coefficients $f_0$ to $f_4$ take the values
\begin{eqnarray}
f_{0}&=&-\frac{{\MP}^{2}}{16\pi},
\nonumber
\\
f_{1}&=&
a_{1} + a_{2} + \frac{1}{2 \sqrt{-b}}
\,\big( \omega \psi - \omega \varphi  - b\varphi \big) 
\nonumber
\\
f_{2}&=&
-2a _{1} - 4a_{2}
+ \frac{1}{\sqrt{-b}}
\,\big(\omega \varphi  + 2b\varphi  -  \omega \psi \big),
\nonumber
\\
f_{3}
&=&
\frac{a _{1}}{3} + a_{2}  - \frac{3c + 2b}{36}
+ \frac{1}{6\sqrt{-b}}
\,\big(\omega \psi  - \omega \varphi  - 3b\varphi\big),
\nonumber
\\
f_{4}
&=&
-\frac{4\pi}{3}\,\sqrt{-b},
\label{fs}
\end{eqnarray}
with $ a_{1,2}$ defined through (\ref{vacuum}) and we have neglected the
cosmological constant since we shall be concerned with the high energy
branch of the solution \eqref{H}. Note actually that although the
combinations here presented are synthetic and exhaustive, the actual
equations of motion derived from \eqref{1} do not depend on $a_2$ since
it comes from a surface term.
	
Using the notation $h_{i}^{\  j}\to h$, where $h$ now stands for a
tracefree tensor perturbation, one obtains the perturbation equation
\cite{Fabris:2011qq}, which reads
\begin{eqnarray}
0
&=&
\Big( 2 f_{1} + \frac{f_{2}}{2}\Big)  h^{\lp\textsc{\tiny IV}\rp}
+ \Big[ 3H(4f_1 + f_2) + 4 \dot{f_1} +
\dot{f_{2}}\Big]  h^{\lp\textsc{\tiny III}\rp}
+ \Bigl[ 3H^{2}\Big(6f_{1} + \frac{f_2}{2} - 4f_{3}\Big)
\nonumber
\\
\nonumber
&+& H\Big(16 \dot{f_{1}} + \frac{9}{2}\dot{f_{2}}\Big)
+ 6\dot{H}(f_1 - f_3) + 2\ddot{f_{1}}
+ \frac{1}{2}\big(\ddot{f_2} + f_0 + f_4\ddot{\varphi}\big)
+ \frac{3}{2} f_{4} H \dot{\varphi}
- \frac{1}{3} \dot{\varphi}^{2}\Bigl] \ddot{h}
\\
\nonumber
&-& (4 f_{1} + f_{2}) \frac{\nabla^2 \ddot{h}}{a^2}
+ \Big[ 2\dot{H}(2\dot{f_{1}} - 3 \dot{f_{3}})
- \frac{21}{2}\, H \dot{H} \Big(f_2 + 4 f_3\Big)
- \frac32\,\ddot{H}\Big(f_2 + 4 f_3\Big)
\\
\nonumber
&+& 
3H^{2}\Big(4 \dot{f_{1}} + \frac{1}{2} \dot{f_{2}}
- 4 \dot{f_{3}}\Big)
-9H^{3}\big(f_2 + 4f_3\big)
+ H \Big( 4\ddot{f_{1}} + \frac{3}{2}\ddot{f_{2}}\Big)
+ \frac{3}{2} f_4 \dot{\varphi}(3 H^{2} + \dot{H})
\\
\nonumber
&+& 
H \Big( 3f_4 \ddot{\varphi} + \frac{3}{2} f_0
- \dot{\varphi}^2 \Big)
+ \frac{1}{2} f_{4} \stackrel{...}{\varphi}
- \frac{2}{3} \dot{\varphi} \ddot{\varphi} \Bigl]  \dot{h}
- \big[H(4 f_{1} + f_{2}) + 4 \dot{f_{1}}
+ \dot{f_{2}}\big] \frac{\nabla^{2} \dot{h}}{a^{2}}
\\
\nonumber
&+& \Big[5 f_{4} H \stackrel{...}{\varphi}
+ f_{4} \stackrel{....}{\varphi}
- \big(36 \dot{H} H^{2}
+ 18 \dot{H}^{2}
+ 24 H \ddot{H}
+ 4\stackrel{...}{H}\big) (f_{1} + f_{2} + 3 f_{3})
\nonumber
\\
\nonumber
&-& 
4H \dot{H}(8 \dot{f_{1}} + 9 \dot{f_{2}} +
30 \dot{f_{3}})
- 8\ddot{H}(\dot{f_{1}} + \dot{f_{2}} + 3 \dot{f_{3}})
- H^{2}(4 \ddot{f_1} + 6 \ddot{f_2} + 24 \ddot{f_3})
\\
\nonumber
&-& 
4\dot{H}(\ddot{f_1} + \ddot{f_2} + 3 \ddot{f_3})
- 9f_{4}\dot{\varphi}(H^3 + H \dot{H})
+ f_{4} \ddot{\varphi}(3 H^{2} + 5 \dot{H})
- H^{3}(8 \dot{f_{1}} + 12 \dot{f_{2}} + 48 \dot{f_{3}})
\\
\nonumber
&+&
\frac{1}{12} \dot{\varphi}^{2}\big(3H^{2} + 2\dot{H}\big)
 + \frac{1}{3} H \dot{\varphi} \ddot{\varphi}
- \frac{1}{12} \ddot{\varphi}^{2}
+ \frac{1}{6} \dot{\varphi} \stackrel{...}{\varphi}
+ f_{0}(2 \dot{H} + 3 H^{2})\Bigl] h
\\
\nonumber
&+&
\Bigl[
2\big(2H^2 + \dot{H}\big)(f_{1} + f_{2} + 3 f_{3})
+ \frac{1}{2} H \big(4 \dot{f_{1}} +  \dot{f_{2}}\big)
\\
&-& \frac{1}{2}  \big(\ddot{f}_2
+ f_4 \ddot{\varphi} + f_0  + 3 f_4 H \dot{\varphi} \big)
- \frac{1}{6} \dot{\varphi}^{2}\Bigl] \frac{\nabla^{2} h}{a^{2}}
+ \Bigl(2 f_{1}
+ \frac{1}{2} f_{2}\Bigl) \frac{\nabla^{4} h}{a^{4}}.
\label{diff}
\end{eqnarray}
If we take $H$ to be constant in \eqref{diff}, we recover the result of
Ref.~\cite{Gasperini:1997up}. We also check explicitly that $a_2$
cancels off systematically from all the coefficients appearing in this
equation.

Eq.~\eqref{diff} describing tensor perturbation dynamics with quantum
anomaly-induced corrections is rather cumbersome, and
Ref.~\cite{Fabris:2011qq} also provides all the missing details that may
happen to be necessary. For the purpose of illustration, we also separate 
the relevant equation without the quantum terms, which is much simpler
\cite{Shapiro:2014fsa}. It reads
\begin{eqnarray}
\frac{1}{3}  h^{\lp\textsc{\tiny IV}\rp}
&+&
2H h^{\lp\textsc{\tiny III}\rp}
+ \lp H^2 + \frac{\MP^2}{32\pi a_1}\rp \ddot{h}
+ \frac{2}{3} \lp \frac14 \frac{\nabla^4 h}{a^{4}}
- \frac{\nabla^2 \ddot{h}}{a^2}
-  H \frac{\nabla^{2} \dot{h}}{a^{2}}\rp
\nonumber \\
&-&
\lp H \dot{H} + \ddot{H} + 6 H^3
 - \frac{3\MP^2 H}{32\pi a_1}\rp \stackrel{.}{h}
 - \lc \frac{\MP^2}{32\pi a_1}
- \frac43  \lp \dot{H} + 2H^2\rp
\rc \frac{\nabla^2 h}{a^{2}}
\nonumber \\
&-&
\lc
24 \dot{H} H^{2} + 12 \dot{H}^{2} + 16 H \ddot{H}
+ \frac83 H^{\lp\textsc{\tiny III}\rp}
-\frac{\MP^{2}}{16\pi a_1}\lp 2 \dot{H} + 3 H^{2}\rp\rc h
 = 0,
\label{pertu}
\end{eqnarray}
and depends only on the coefficient $a_1$ in the action (\ref{HD}). 
As discussed in
\cite{Fabris:2011qq}, for frequencies much below the Planck scale, the
linear stability analysis should be expected to give the same results in the 
cases of Eq.~\eqref{pertu} and Eq.~(\ref{diff}), that is for the complete 
theory with anomaly-induced terms. Since we shall indeed consider 
Planck frequencies, the complete equation \eqref{diff} is the relevant 
one; it is this equation we use in what follows, but in most cases simplified 
slightly by assuming a de Sitter background, i.e. by setting $\dot H\to 0$.
\end{widetext}

\begin{figure}[ht]
	\centering
	\includegraphics[scale=0.35]{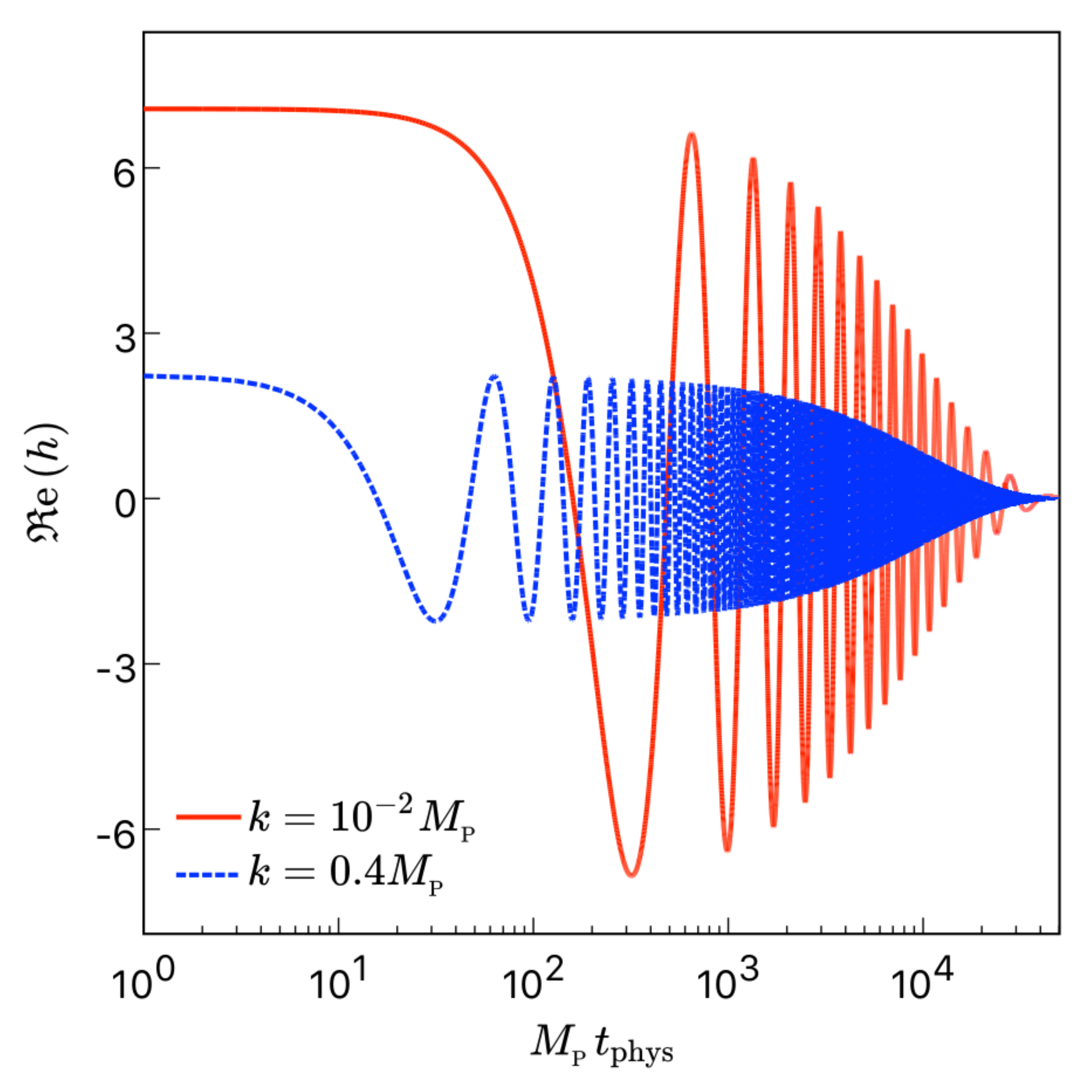}
	\caption{Real part of the gravitational mode amplitude, solution of
		the simplified Eq.~\eqref{pertu}, as a function
		of time, rescaled in units of the Planck time $\MP^{-1}$ for two values of
		the wavenumber, namely $k=10^{-2}\MP$ (full line) and $k=0.4 \MP$
		(dashed line) and a Hubble constant set to $H=10^{-4}\MP$. Here and
		in the following, we set $a_1 = -1$.}
	\label{HDcase}
\end{figure}

\section{Initial conditions}

From an observational point of view, the last missing piece of the
primordial puzzle, also an important characteristics of inflationary
predictions, is the tensor spectrum, i.e. the scale distribution of
gravitational waves produced during the quasi de Sitter phase. As in the
more traditional models, in the context of anomaly-induced quantum
gravity, both the scalar and tensor perturbations are generated as
quantum vacuum fluctuations subsequently freezing out at Hubble
crossing. In what follows, we accordingly set the initial conditions of
the tensor mode by assuming it begins in the Bunch-Davies vacuum state;
more details about these initial conditions can be found in
\cite{Grishchuk:1993te}) and in \cite{Fabris:2011qq} for the 
case with anomaly-induced corrections.

The procedure we assume consists in first quantizing the tensor
perturbations in the usual way, by merely considering the
Einstein-Hilbert action expanded to second order. The canonical 
creation and annihilation operator expansion then allows to set the 
vacuum initial condition for the field, which we then generalize to 
the higher derivative and the anomaly-induced terms.

\subsection{The usual Einstein-Hilbert case}

Substituting Eq.~(\ref{pertmet}) into the Einstein-Hilbert action
\eqref{EH}, one finds
\begin{equation}
S_h=\frac12 \MP^2\int \dd\eta\,\dd^3x \,a^2(\eta)
\lc \lp h^{\prime}_{ij}\rp^2 - \lp \nabla_k h_{ij}\rp^2\rc,
\label{EHpert}
\end{equation}
where a prime denotes a derivative with respect to the conformal time
$\eta$. A canonical field formulation is obtained through the following
time normalization
\begin{equation}
h_{ij}(\bm{x},\eta)=\frac{\mu_{ij}(\bm{x},\eta)}{a(\eta) \MP},
\end{equation}
thereby defining $\mu_{ij}$. This transforms Eq.~(\ref{EHpert}), up to
an irrelevant time derivative, into
\begin{equation}
S_\mu = \frac12 \int \dd\eta\,\dd^3x\, a^2(\eta)\Big[ 
\lp \mu^{\prime}_{ij}\rp^2
- \lp\nabla_k \mu_{ij}\rp^2 +
\frac{a^{\prime\prime}}{a}\mu^{2}_{ij}\Big] ,
\label{Smu}
\end{equation}
which corresponds to the action for a scalar field with time-dependent
mass. We further decompose $\mu_{ij}$ into two independent modes in
Fourier space as
\begin{widetext}
\begin{equation}
\mu_{ij}(\bm{x},\eta) = \int \frac{\dd^3 \bm{k}}{(2\pi)^32 k}
\ex^{i \bm{k}\cdot \bm{x}}
\sum_{\lambda=+,\times} \left[\hat{a}^{\lambda}_{\bm{k}}
\mu^{\lambda}_{k}(\eta) e^{\lambda}_{ij}(\bm{k}) +
\hat{a}^{\lambda\dag}_{-\bm{k}}\mu^{\lambda*}_{k}(\eta)
e^{\lambda*}_{ij}(-\bm{k})\right],
\label{Fourier}
\end{equation}
\end{widetext}
where we have introduced the polarization tensors
$e^{\lambda}_{ij}(\bm{k})$, satisfying the symmetric
[$e^{\lambda}_{ij}(\bm{k})= e^{\lambda}_{ji}(\bm{k})$], transverse
[$k_i e^{\lambda}_{ij} (\bm{k})=0$], traceless
[$e^{\lambda}_{ii}(\bm{k})=0$] and orthogonality
[$e^{\lambda}_{ij}(\bm{k}) e^{\lambda^{\prime}}_{ij}
(\bm{k})=\delta^{\lambda\lambda^{\prime}}$] conditions (see
\cite{Peter:2013woa} for details on the structure of these tensors).

\begin{figure*}[ht]
	\centering
	\includegraphics[scale=0.35]{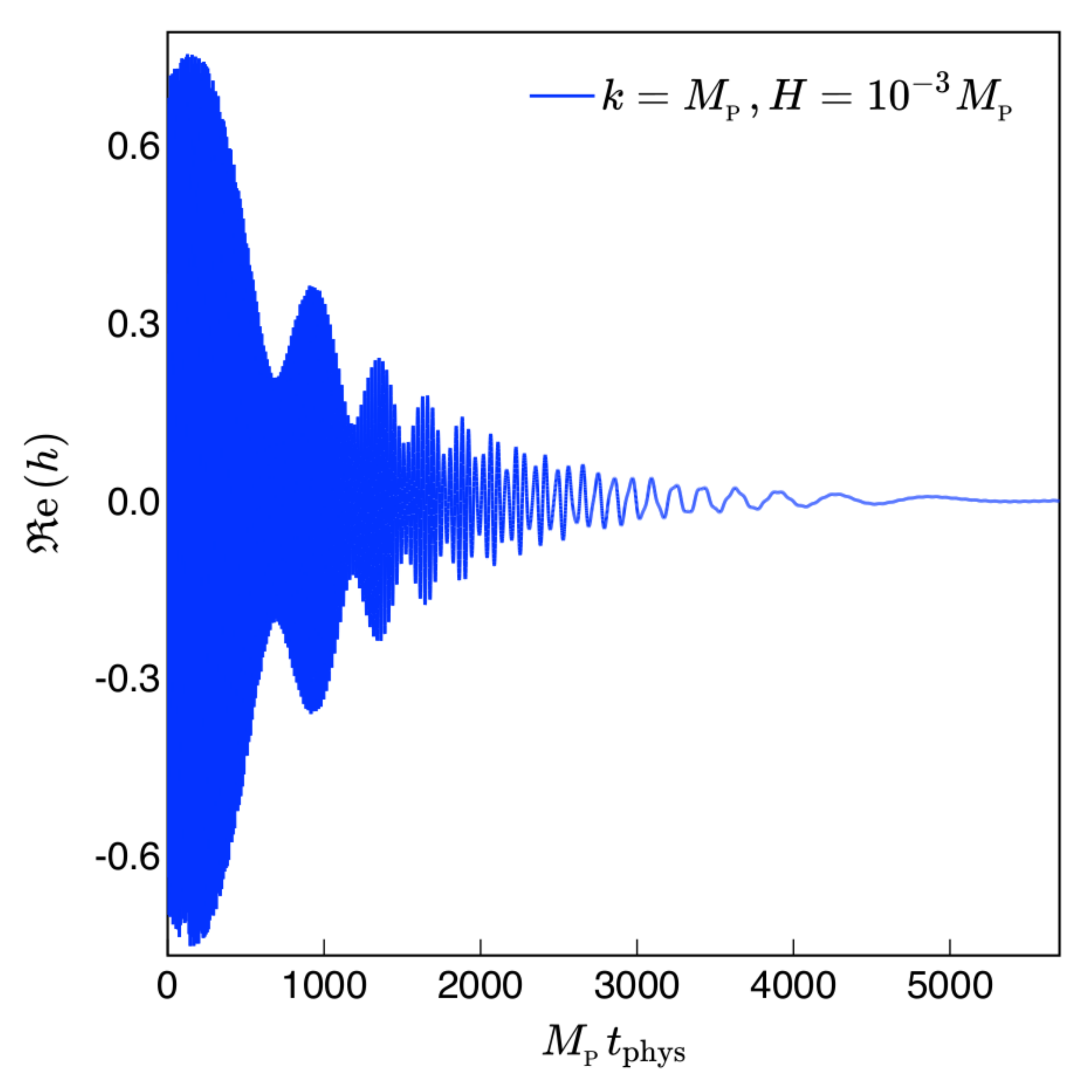}
	\includegraphics[scale=0.35]{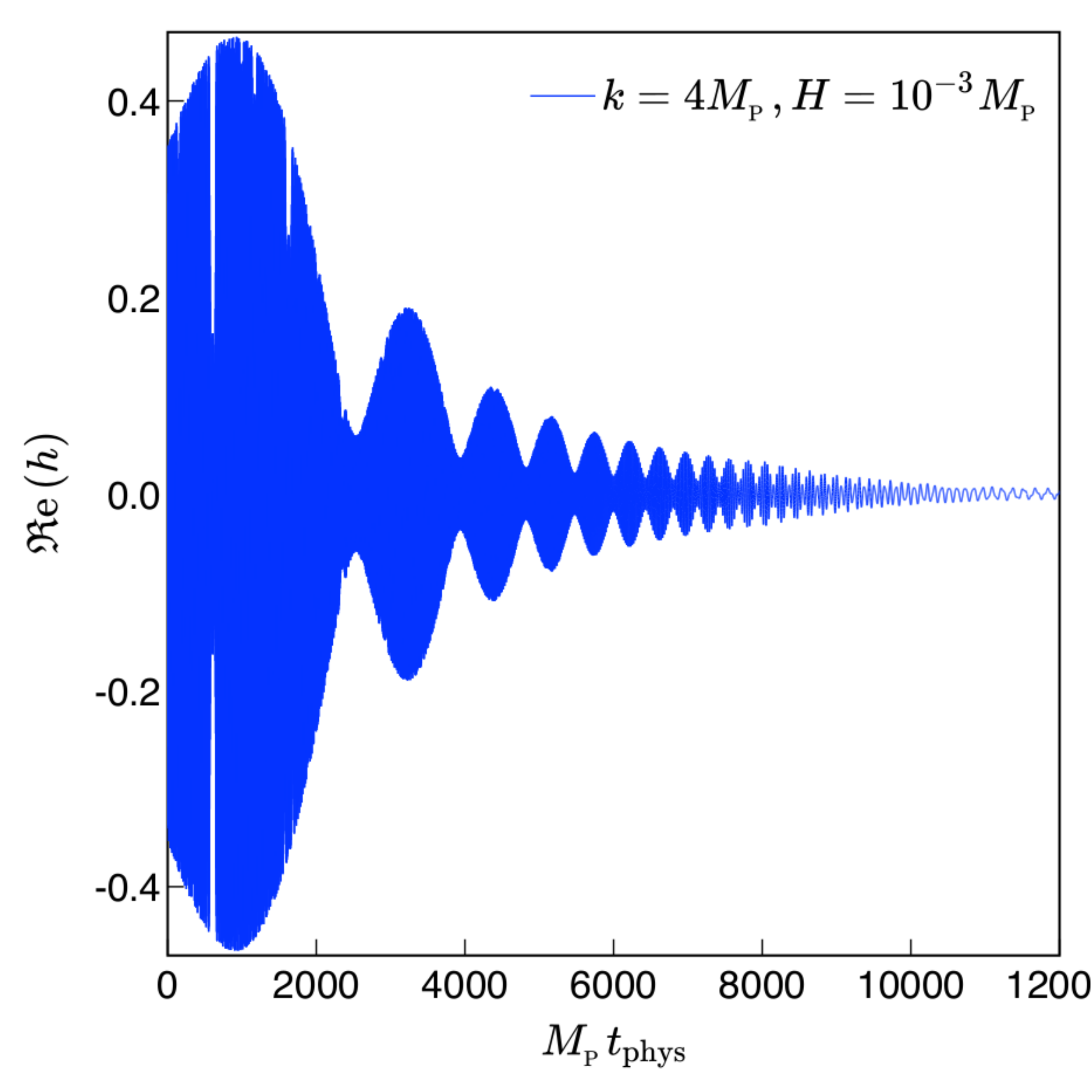}
	\includegraphics[scale=0.35]{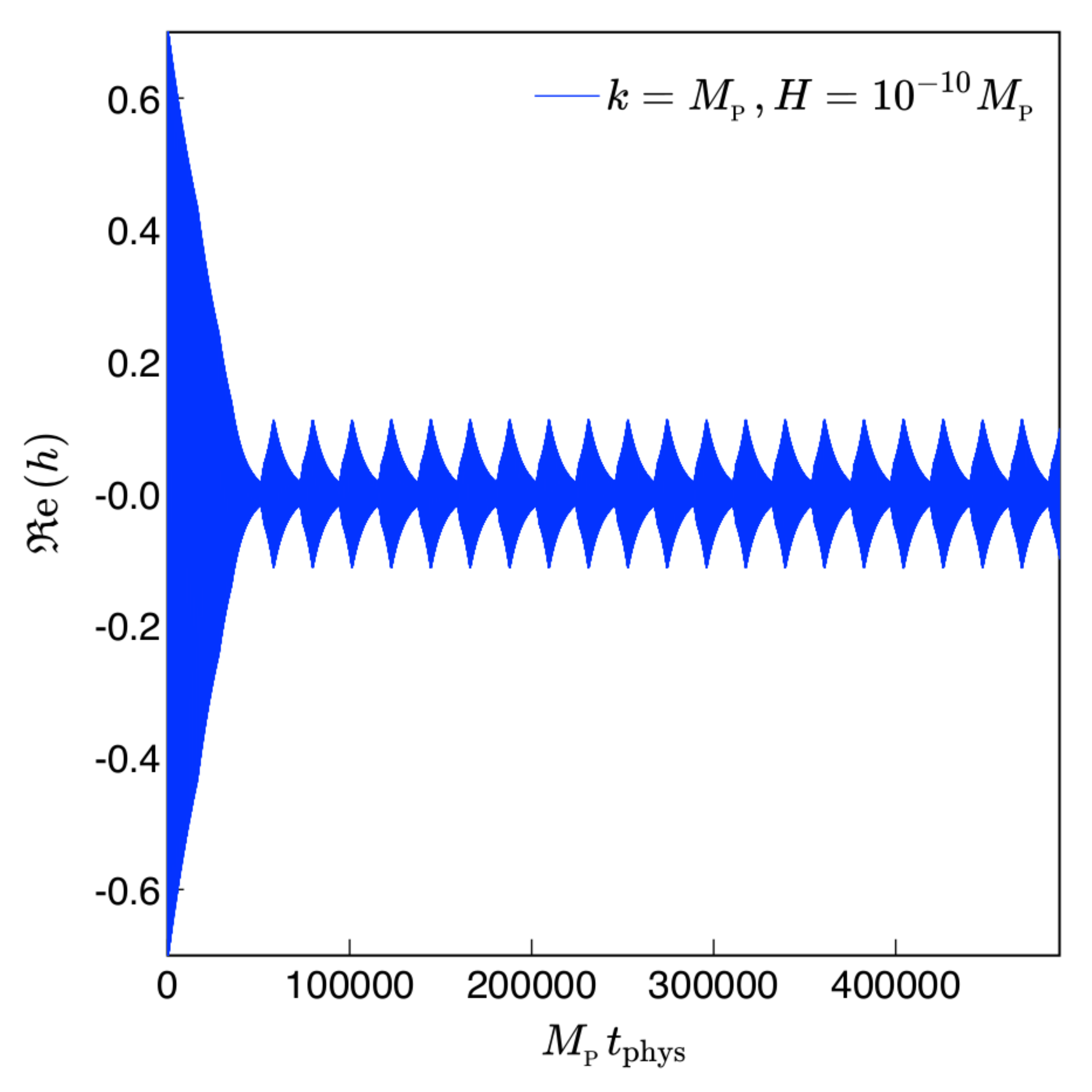}
	\includegraphics[scale=0.35]{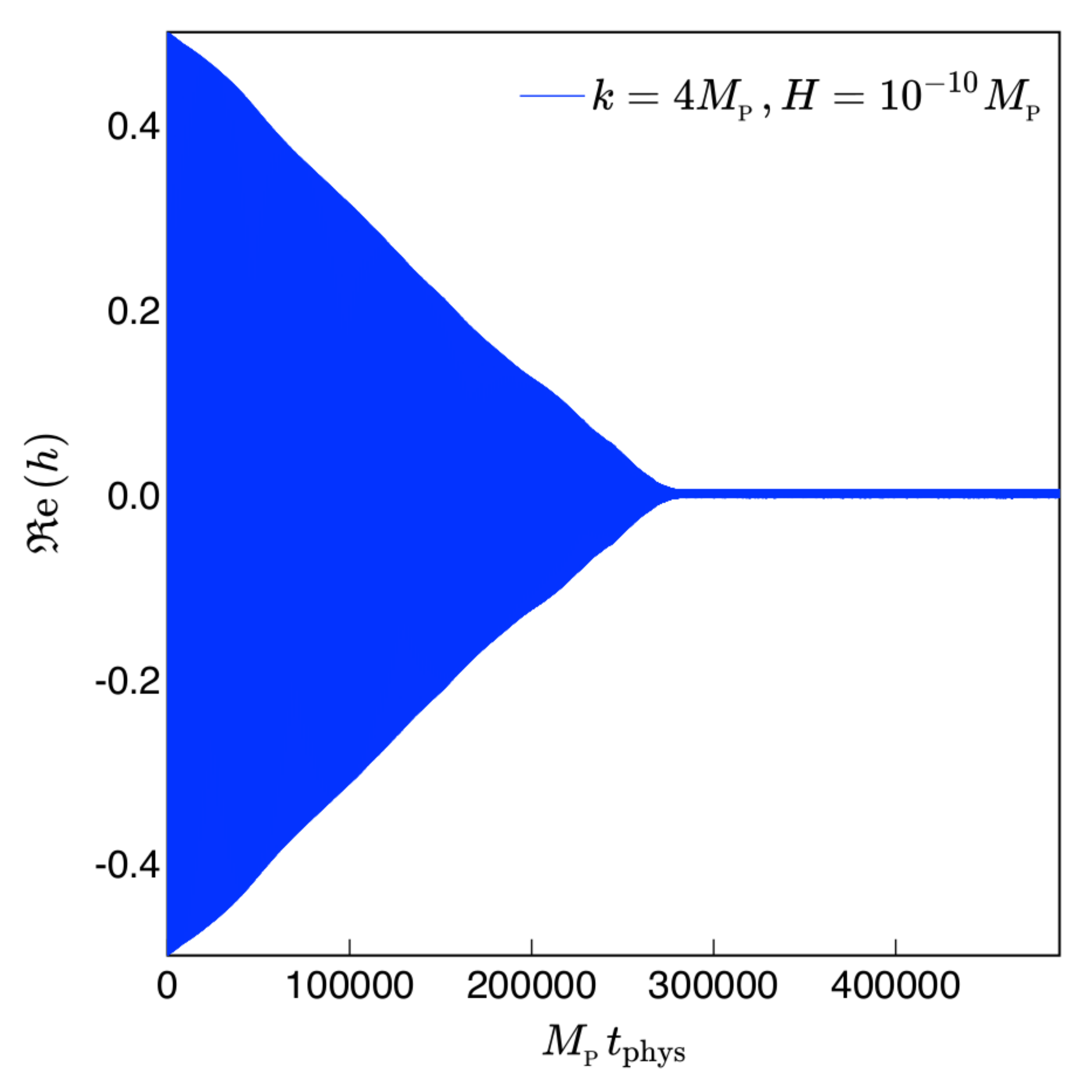}
	\caption{Real part of the gravitational mode solution of the full
		equation \eqref{diff} for two distinct de Sitter cases, high (top
		panel) and low (bottom panel) Hubble rates $H=10^{-3}$ and $H=10^{-10}$
		(in Planck units) respectively. Each case shows two Planckian modes,
		namely $k=1$ (left panel) and $k=4$ (right panel). All plots are for
		$a_1 = -1$. The top panels shows that even though the initially
		increasing mode oscillate tremendously, the damping due to the very
		rapid background together with the redshifting of the mode wavelength
		itself manage to suppress the mode amplitude. No runaway solution is
		found in such a case. The bottom panels on the other hand show that the
		mode rapidly decreases and is subsequently halted to a constant
		amplitude due to the fact that the Hubble rate is not large enough to
		fully compensate for the runaway.}
	\label{pert_est}
\end{figure*}

Since Eq.~\eqref{EHpert} is linear, the expansion \eqref{Fourier}
plugged back into \eqref{Smu} yields independent contributions from both
$+$ and $\times$ modes for each value of the wavenumber $k$. Quantizing
can be done on each polarization mode separately as if those were
independent scalar fields: the gravitational wave part of the action at
second order is nothing but a sum over independent parametric
oscillators with time-dependent frequencies, for which the standard
rules apply. In particular, one can impose the usual canonical
commutation relations on the creation and annihilation operators
$\hat{a}^{\lambda}_{\bm{k}}$, $\hat{a}^{\lambda\dag}_{\bm{k}}$. They
read
\begin{eqnarray}
\left[\hat{a}^\lambda_{\bm{k}},
\hat{a}^{\lambda'\dag}_{\bm{k}^\prime}\right] &=& (2\pi)^3
\delta^{\lambda\lambda^{\prime}} \delta(\bm{k} - \bm{k}^\prime) , \\
\left[\hat{a}^{\lambda}_{\bm{k}},
\hat{a}^{\lambda^{\prime}}_{\bm{k}^\prime}\right] &=& 0 , \\
\left[\hat{a}^{\lambda\dag}_{\bm{k}},
\hat{a}^{\lambda^{\prime}\dag}_{\bm{k}^\prime}\right] &=& 0 .
\end{eqnarray}

The normalization condition of the mode function $\mu_k(\eta)$ is given
by that of the Wronskian
\begin{equation}
\mu^{\lambda*}_{k} \partial_\eta \mu^{\lambda}_{k} -
\mu^{\lambda}_{k} \partial_\eta \mu^{\lambda*}_{k} = i,
\end{equation}
and from the action~(\ref{Smu}), we find that the equation of motion for
a given mode function, irrespective of the polarization (which is then
dropped out of the equation), is given by
\begin{equation}
\mu^{\prime\prime}_{k} + \left(k^2 -
\frac{a^{\prime\prime}}{a}\right)\mu_k = 0.
\label{grcase}
\end{equation}
As argued above, this is indeed the (or more accurately two copies of
the) equation for a parametric oscillator with time-dependent frequency
\cite{Mukhanov:1990me,Peter:2013woa,PPJPU2013}.

We are now in a position to impose the Bunch-Davies initial conditions
for the modes discussed above. For the sub-Hubble modes ($k\gg aH$), one
gets the oscillatory behavior typical of the Minkowski vacuum,
\begin{equation}
k\gg aH \ \Longrightarrow \  \mu^{\prime\prime}_k + k^2 \mu_k = 0
\,\Longrightarrow\,    \mu_k =
\frac{\ex^{-i k \eta}}{\sqrt{2 k}},
\end{equation}
while on the other hand, on super-Hubble scales, the modes scales with
the scale factor, namely
\begin{equation}
k\ll aH \ \Longrightarrow \  \mu^{\prime\prime}_k - \frac{a^{\prime\prime}}{a}
\mu_k = 0
\,\Longrightarrow\,   \mu_k \propto a.
\end{equation}
From the mode equation \eqref{grcase}, the solution changes from
oscillatory to growing at Hubble crossing $k\sim aH$.

\subsection{Seeding the perturbations in the general case}

We are now in a position to solve either \eqref{diff} for the general
case or \eqref{pertu}, restricting attention as a first approximation to
this simpler case. Although the variable $\mu_k(t)$ is extraordinary
useful to build the actual observational tensor spectrum, we shall only
make use of it to provide relevant initial conditions when the extra
terms leading to the fourth derivative equation of motion are small.

Let us first expand the actual quantity of interest here, namely the
matrix $h(\bm{x},t)$, which we expand in Fourier modes just like its
standard counterpart $\mu_k(t)$. Assuming fluctuations of the zero point
energy of the latter, we find that we can consider an independent
mode by writing
\begin{equation}
h(\bm{x},\eta) = h(\eta)\ex^{\pm i\bm{k}\cdot\bm{x}}
\ \ \ \hbox{and} \ \ \
h(\eta) = \frac{\ex^{\pm i k\eta}}{a\MP\sqrt{2k}}.
\end{equation}
We wrote the last expressions in terms of the conformal time, since it
is this time that renders the FLRW metric conformal to the Minkowski
metric in flat space; $k$ is the comoving wavenumber vector. After
fixing the initial value of the mode $\mu$ and hence $h$, it is a simple
matter to derive how the initial amplitude depends on $k$. In our case,
in the limit $k\gg aH$, this simplifies to
\begin{equation}
h_\mathrm{ini}\varpropto\frac{1}{\sqrt{2k}},\ \
\dot{h}_\mathrm{ini}\varpropto\sqrt{\frac{k}{2}},\ \
\ddot{h}_\mathrm{ini}\varpropto\frac{k^{3/2}}{\sqrt{2}},\ \
h^{\lp\textsc{\tiny III}\rp}_\mathrm{ini}
\varpropto\frac{k^{5/2}}{\sqrt{2}}.
\label{initcond}
\end{equation}
Although the initial conditions are set in terms of the conformal time
$\eta$, we translate those in terms of the cosmic time $t$ since it is
the latter rather than the former which is used in Eqs.~\eqref{diff} and
\eqref{pertu}. Up to some irrelevant constants of order unity (normalizing
the scale factor to one at the initial time) and the Planck mass $\MP$,
Eq.~\eqref{initcond} provides the relevant initial conditions for
solving the cases of interest.

\section{Evolution}

We will now rewrite the terms in (\ref{pertu}) or \eqref{diff} using the
standard Fourier transform in the space variable, namely we assume the
replacement $\bm{\nabla} \to i \bm{k}$ (i.e. $\nabla^2 \to -k^2$).
The gravitational waves we are interested in appear during the inflation
phase because of the quantum gravitational fluctuations, i.e., the
generation of initial seeds of primordial gravity waves by inflation is
a quantum process, while their further dynamics can be explored as a
classical phenomenon, solving the dynamical equations with quantum
initial conditions. We now want to investigate how the perturbation
amplitudes in the primordial universe depends of the initial frequency
and the background metric dynamics.

For the numerical results presented below, and because the perturbation
equations are linear, we have set $\MP\to 1$, so that both the amplitude
$h$ and the wavenumber $k$ are pure numbers, assumed in units of $\MP$.
The strategy we adopted consists in the following: first, we consider
the simplified version of the mode equation \eqref{pertu} in a de Sitter
case, assuming the Hubble rate to be given independently. This is
possible since we are interested in the tensor modes, and those do not
affect the background evolution at this order. We then move on to the
full equation \eqref{diff} using the MSSM underlying parameters. We also
chose to fix the unknown parameter $a_1$ to $a^\mathrm{num}_1=-1$ as it
should be negative to avoid tachyonic ghosts; it is subsequently
``normalized'' for representational convenience.

\begin{figure*}[ht]
	\centering
	\includegraphics[scale=0.35]{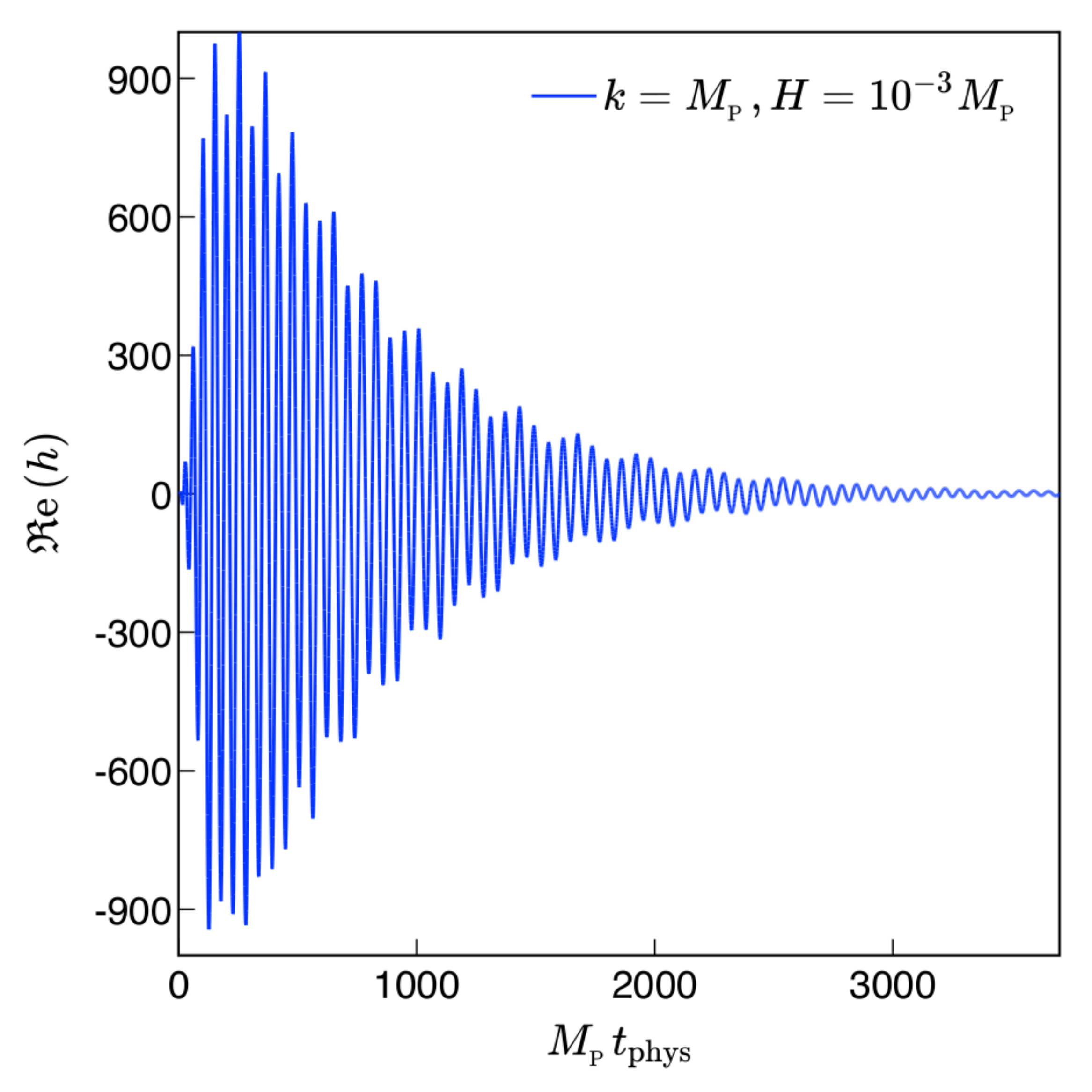}
	\includegraphics[scale=0.35]{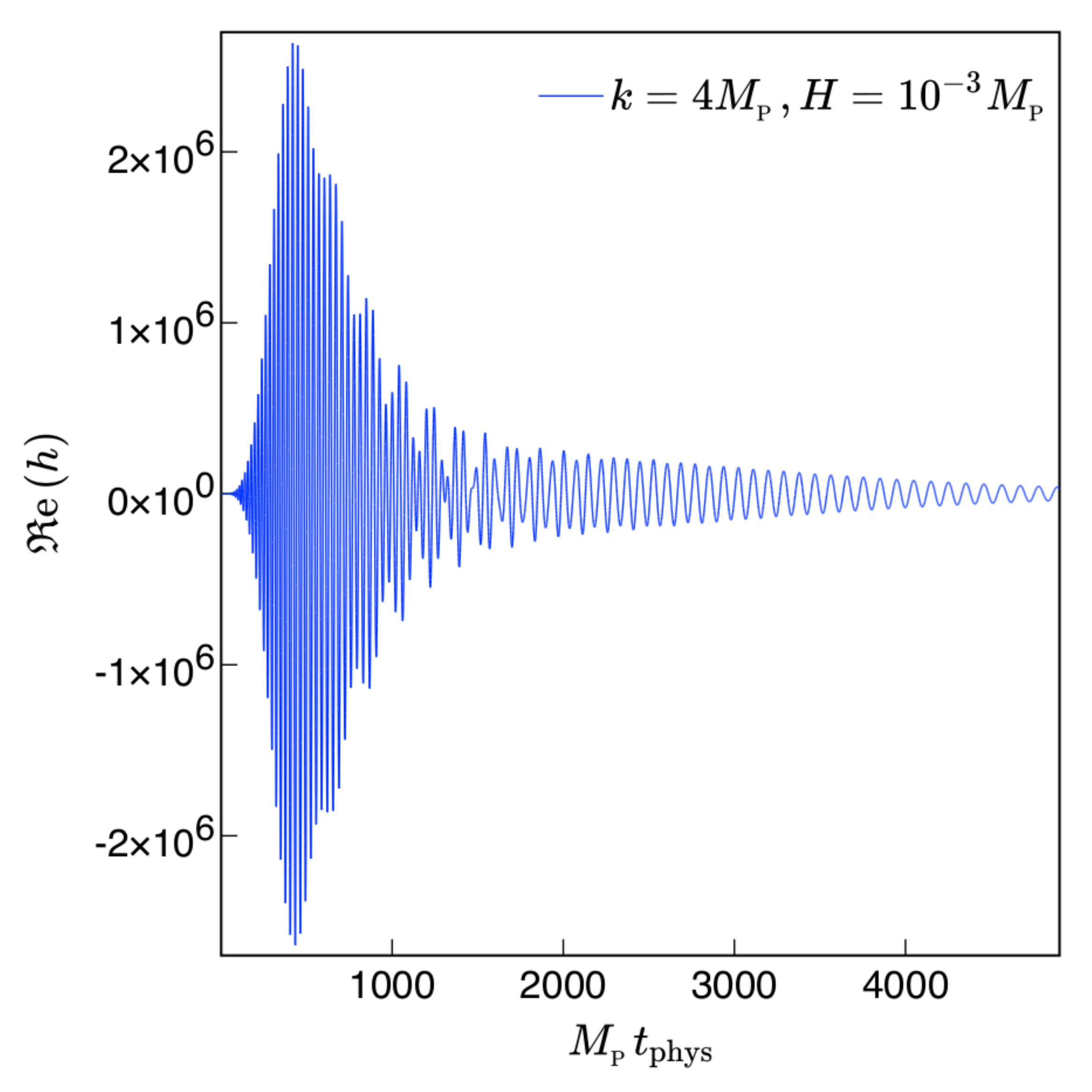}
	\includegraphics[scale=0.35]{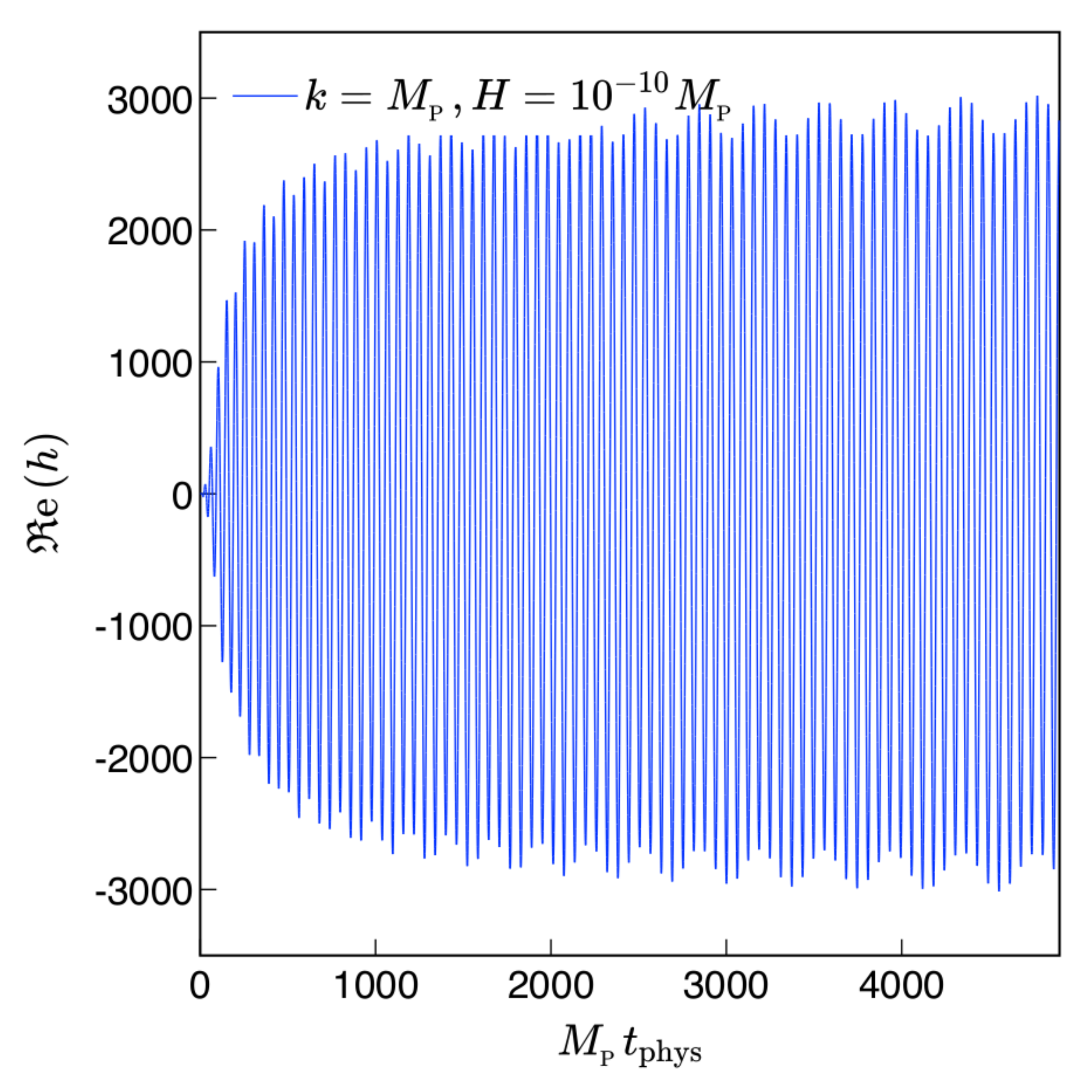}
	\includegraphics[scale=0.35]{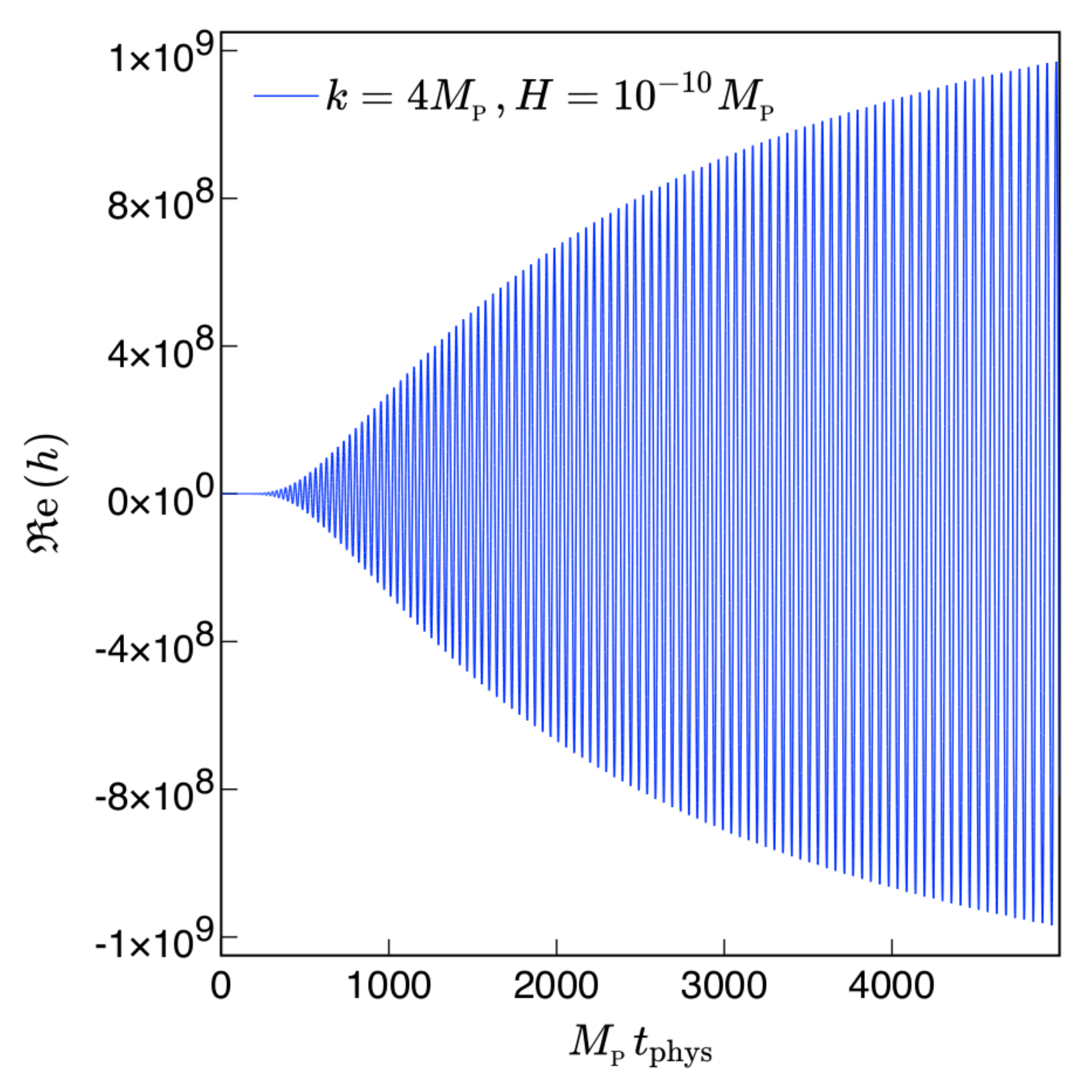}
	\caption{Same as Fig.\ref{pert_est} with a radiation dominated
		background with scale factor behavior $a\varpropto t^{1/2}$. Again, the
		top panels are for a large initial Hubble scale (now time-varying)
		$H=10^{-3}$ and the bottom panels for low initial Hubble rate
		$H=10^{-10}$ (still in Planck units) and the two relevant Planckian
		modes $k=1$ and $k=4$ are displayed on the left and right panels
		respectively. It is clear from the figure that the case modes, although
		initially setting themselves on the runaway ghost solutions, are
		eventually damped thanks to the large initial value of the Hubble
		scale, which manages to take up the mode evolution. The growth is
		however uncontrolled when the initial expansion rate is too low, so the
		background can keep up with the runaway.}
	\label{pert_est1}
\end{figure*}

\subsection{Higher derivative correction}

When one considers only general relativity, i.e. using only
$S_{_\mathrm{EH}}$ of Eq.~\eqref{EH}, the initial frequency
$k$ hardly changes anything at all in the evolution of the mode
except though its normalization. The theory being stable, we
find, as expected, so ghost solution whatever the value of either
the wavelength $k$ or the Hubble parameter $H$, even for a
value as low as the present-day estimate $H=10^{-61}$
and $k$ of order unity.

The vacuum action $S_\mathrm{vac}$ of Eq.~\eqref{vacuum} yields,
in practice, a fourth-order differential equation subject to plausible
instabilities. Provided $k\ll 1$, one expects to describe the full situation
by merely considering this simplified version.

Fig.~\ref{HDcase} exemplifies a situation where the Hubble rate is
in a reasonable range in which one expects the corrections to have
non negligible effects while at the same time not demanding a full
quantum gravity theory. In practice, we set $H=10^{-4}$. Such a
high value permits to control the unwanted runaway solution: even
for rather high wavenumbers, the relevant term in Eq.~\eqref{pertu}
is proportional to $\nabla^2 h/a^2\to k^2 h /a^2$, and with $a\sim
\ex^{Ht}$, this contribution goes exponentially fast to negligible values
provided the background Hubble rate is sufficiently large.

On the other hand, starting with a trans-Planckian 
wavelength $k>1$ immediately leads to the ghost taking over the 
background, and  the runaway solution is not possible to stop, at 
least for reasonable values of the Hubble rate. In order to increase 
this Hubble rate and figure out the consequence of such an increase, 
one needs to use the full equation \eqref{diff} including the 
anomaly-induced quantum corrections, to which we now turn.

\subsection{Anomaly-induced contribution}

The same $(k/a)^2$ term in Eq.~\eqref{pertu} is present in the full
version \eqref{diff}, which also contains an additional $(k/a)^4$. By
the same reasoning, if the expansion is fast enough, these terms 
rapidly become negligible and one should expect the trans-Planckian
runaway to become a non-issue, given a strong enough, sufficiently
fast-evolving, background. The case $k<1$ can be studied through the
simplified version of \eqref{pertu}, which has shown the ghost
instability to be essentially irrelevant. It remains to be seen what
happens in the more complicated situation where the mode is
trans-Planckian, and for this we now discuss the relevant solutions of
Eq.~\eqref{diff}.

In a first analysis, we studied the de Sitter case under two separate
assumptions, namely that the mode is exactly Planckian, i.e., $k=1$, and
we then considered the extreme situation with $k>1$, and we set it
arbitrarily to $k=4$. For both these modes, we considered either a slow
expansion with $H=10^{-10}$, and a fast one with $H=10^{-3}$.
Fig.~\ref{pert_est} presents these solutions which, we must add, are
representative of the general solution and are merely meant to
exemplify the underlying results.

We note that for a de Sitter background with $a\varpropto \ex^{Ht}$ and
$H$ constant, a high value of $H$ induces the following behavior for the
mode functions: first, it seems to increase in amplitude, seemingly
initiating the ghost instability. Then, the background expanding very
rapidly, the $(k/a)^2$ and $(k/a)^4$ terms are damped to become
vanishingly small, and quickly negligible compared to the other terms of
Eq.~\eqref{diff}. At this point, the mode wavelength has been redshifted
sufficiently that it is no longer trans-Planckian and the usual evolution of
the previous section takes over. The ghost instability is therefore tamed
in this context.

The situation is different when the Hubble rate is much smaller, as
exemplified with the case $H=10^{-10}$ and the same mode numbers. In
this case, the redshifting is not efficient enough, and although the
anomaly-induced terms tend to decrease the amplitude, the effect
eventually saturates and the amplitude ends up being constant.

In both cases, one sees the amplitude is slightly smaller for larger initial
values of $k$, but this is merely due to the vacuum normalization translated
into \eqref{initcond}.

Finally, the situation is even more interesting in the case of a
radiation dominated universe for which we fix the initial value of the
Hubble rate. Again, we find, as shown in Fig.~\ref{pert_est1}, that an
initially large expansion rate eventually catches up with the ghost
runaway and drives it back to small values. It is unclear however that
this could not trigger higher order perturbations and possibly
backreaction; this point deserves further examination.

Lastly, a radiation dominated universe with low expansion rate and
trans-Planckian modes is definitely problematic as in all cases studied,
exemplified in the bottom panels of Fig.~\ref{pert_est1}, as the runaway
solution seems to increase unboundedly, with the higher wavenumber
implying a larger divergence. At this stage, there does not seem to
exist any foreseeable mechanism susceptible to tame this instability.

\section{Conclusions}

We have studied the anomaly-induced would-be ghost instability
in the framework of a FLRW cosmological background with
tensor perturbations. As found in a previous works
\cite{Fabris:2011qq,Salles:2014rua},
we confirm that sub-Planckian modes on a slowly
varying background do not exhibit runaway solutions. Moreover,
we find that sub-Planckian on a rapidly varying background
also do not exhibit this instability.

The crucial new point concerns the very rapidly expanding background
case. In that situation, we found that even trans-Planckian modes can 
be quickly redshifted and soon become effectively sub-Planckian, so 
that the background acts as a runaway controller. We have thus 
obtained a way to tame this otherwise hardly ever mentioned instability, 
at least in a cosmological context. 
Given the above argument, we are able to conjecture that the ghost
instability may actually be a non-issue in the cosmological setting: if 
trans-Planckian modes can only be produced in a very rapidly varying 
background, indeed one for which the anomaly-induced corrections are
not negligible, then they will be naturally tamed as in our cosmological 
examples. 

It is easy to indicate the main remaining problems on the way to 
better description of the role of ghosts in quantum gravity. First of 
all, it would be very important to study possible realizations of the 
hypothetic mechanism producing an upper bound on the graviton 
density \cite{Dvali:2010ue} or alike, both in general and at least 
on the cosmological 
background. Second, it would be great to explore the possibility to 
suppress the growth of tensor modes on other metric backgrounds, 
capable to develop singularities or at least Planck-order densities 
of the gravitational field. The existing works on this subject 
\cite{Whitt:1985ki,Myung:2013cna,Mauro:2015waa} are not 
conclusive and also so not take into account the non-localities. 
Looking from the general viewpoint, the same mechanism preventing 
a high density of gravitons \cite{Dvali:2011aa} should work in this 
case. It might happen that the problem can be solved by better 
understanding non-localities, as it was originally suggested in 
\cite{Tseytlin:1995uq}.

As a final point, we should like to mention that the instabilities that
may be generated by the ghosts discussed in this paper potentially arise
only at the end of inflation. At this epoch, the non-linear effects are
very likely to end up dominating the overall evolution, and thus are
expected to modify the physical situation drastically. Therefore, linear
perturbation stability cannot, in such a framework, be imposed as an
important condition for a consistent theory of gravity.

\begin{acknowledgments}
F.S. is grateful to CAPES for partial support and CNPq
(grant number 233346/2014-7) for supporting his research project and
especially the visit to IAP/Paris, where an essential part of the work
was done. P.P. would like to thank the Labex Institut Lagrange de Paris
(reference ANR-10-LABX-63) part of the Idex SUPER, within which this
work has been partly done. I.Sh. is grateful to CAPES, CNPq, FAPEMIG
and ICTP for partial support of his work.
The charts in this paper were produced with
Super-Mjograph (http://www.mjograph.net/).
\end{acknowledgments}

\bibliographystyle{apsrev4-1}

\end{document}